\input epsf
\input amssym
\input eplain
\input\jobname.intref

\newfam\scrfam
\batchmode\font\tenscr=rsfs10 \errorstopmode
\ifx\tenscr\nullfont
        \message{rsfs script font not available. Replacing with calligraphic.}
        \def\scr{\cal}
\else   
        \font\sevenscr=rsfs7
        \font\fivescr=rsfs5
        \skewchar\tenscr='177 \skewchar\sevenscr='177 \skewchar\fivescr='177
        \textfont\scrfam=\tenscr \scriptfont\scrfam=\sevenscr
        \scriptscriptfont\scrfam=\fivescr
        \def\scr{\fam\scrfam}
        \def\cal{\scr}
\fi
\catcode`\@=11
\newfam\frakfam
\batchmode\font\tenfrak=eufm10 \errorstopmode
\ifx\tenfrak\nullfont
        \message{eufm font not available. Replacing with italic.}
        
\else
	
	\font\sevenfrak=eufm7 \font\fivefrak=eufm5
        
	\textfont\frakfam=\tenfrak
	\scriptfont\frakfam=\sevenfrak \scriptscriptfont\frakfam=\fivefrak
	
\fi
\catcode`\@=\active
\newfam\msbfam
\batchmode\font\twelvemsb=msbm10 scaled\magstep1 \errorstopmode
\ifx\twelvemsb\nullfont\def\Bbb{\bf}

	\message{Blackboard bold not available. Replacing with boldface.}
\else   \catcode`\@=11
        \font\tenmsb=msbm10 \font\sevenmsb=msbm7 \font\fivemsb=msbm5
        \textfont\msbfam=\tenmsb
        \scriptfont\msbfam=\sevenmsb \scriptscriptfont\msbfam=\fivemsb
        \def\Bbb{\relax\expandafter\Bbb@}
        \def\Bbb@#1{{\Bbb@@{#1}}}
        \def\Bbb@@#1{\fam\msbfam\relax#1}
        \catcode`\@=\active

\fi
        \font\fivemi=cmmi5
        \font\sixmi=cmmi6
        \font\eightrm=cmr8              \def\xrm{\eightrm}
        \font\eightbf=cmbx8             \def\xbf{\eightbf}
        \font\eightit=cmti10 at 8pt     \def\xit{\eightit}

        \font\eighttt=cmtt8

        \font\eightcp=cmcsc8    
                      \def\xold{\eighti}
        \font\eightmi=cmmi8
                     \def\xbold{\eightib}
        \font\teni=cmmi10               \def\old{\teni}
        \font\tencp=cmcsc10

        \font\twelvecp=cmcsc10 scaled\magstep1
        
        \font\sixrm=cmr6
        \font\fiverm=cmr5

        \font\eightsy=cmsy8
        \font\sixsy=cmsy6
        \font\eightsl=cmsl8

        \font\sixbf=cmbx6

	 at10pt	
	\font\twelvehelvbold=phvb at12pt
	 at14pt
	\font\sixteenhelvbold=phvb at16pt
	 at16pt



\def\xbold{\xbf}
\def\xold{\xrm}


\def\noblackbox{\overfullrule=0pt}
\noblackbox

\def\eightpoint{
\def\rm{\fam0\eightrm}
\textfont0=\eightrm \scriptfont0=\sixrm \scriptscriptfont0=\fiverm
\textfont1=\eightmi  \scriptfont1=\sixmi  \scriptscriptfont1=\fivemi
\textfont2=\eightsy \scriptfont2=\sixsy \scriptscriptfont2=\fivesy
\textfont3=\tenex   \scriptfont3=\tenex \scriptscriptfont3=\tenex
\textfont\itfam=\eightit \def\it{\fam\itfam\eightit}
\textfont\slfam=\eightsl \def\sl{\fam\slfam\eightsl}
\textfont\ttfam=\eighttt \def\tt{\fam\ttfam\eighttt}
\textfont\bffam=\eightbf \scriptfont\bffam=\sixbf 
                         \scriptscriptfont\bffam=\fivebf
                         \def\bf{\fam\bffam\eightbf}
\normalbaselineskip=10pt}



\newtoks\headtext
\headline={\ifnum\pageno=1\hfill\else
	\ifodd\pageno{\eightcp\the\headtext}{ }\dotfill{ }{\old\folio}
	\else{\old\folio}{ }\dotfill{ }{\eightcp\the\headtext}\fi
	\fi}
\def\makeheadline{\vbox to 0pt{\vss\noindent\the\headline\break
\hbox to\hsize{\hfill}}
        \vskip2\baselineskip}
\newcount\infootnote
\infootnote=0
\newcount\footnotecount
\footnotecount=1
\def\foot#1{\infootnote=1
\footnote{${}^{\the\footnotecount}$}{\vtop{\baselineskip=.75\baselineskip
\advance\hsize by
-\parindent{\eightpoint\rm\hskip-\parindent
#1}\hfill\vskip\parskip}}\infootnote=0\global\advance\footnotecount by
1}
\newcount\refcount
\refcount=1
\newwrite\refwrite
\def\oldsize{\ifnum\infootnote=1\xold\else\old\fi}
\def\ref#1#2{
	\def#1{{{\oldsize\the\refcount}}\ifnum\the\refcount=1\immediate\openout\refwrite=\jobname.refs\fi\immediate\write\refwrite{\item{[{\xold\the\refcount}]} 
	#2\hfill\par\vskip-2pt}\xdef#1{{\noexpand\oldsize\the\refcount}}\global\advance\refcount by 1}
	}
\def\refout{\eightpoint\catcode`\@=11
        \xrm\immediate\closeout\refwrite
        \vskip2\baselineskip
        {\noindent\twelvecp References}\hfill\vskip\baselineskip
        \baselineskip=.75\baselineskip
        \input\jobname.refs
        \baselineskip=4\baselineskip \divide\baselineskip by 3
        \catcode`\@=\active\rm}

\def\skipref#1{\hbox to15pt{\phantom{#1}\hfill}\hskip-15pt}

\def\hepth#1{\href{http://xxx.lanl.gov/abs/hep-th/#1}{arXiv:\allowbreak
hep-th\slash{\xold#1}}}

\def\arxiv#1#2{\href{http://arxiv.org/abs/#1.#2}{arXiv:\allowbreak
{\xold#1}.{\xold#2}}} 
 
\def\jhep#1#2#3#4{\href{http://jhep.sissa.it/stdsearch?paper=#2\%28#3\%29#4}{J. High Energy Phys. {\xbold #1#2} ({\xold#3}) {\xold#4}}}

\def\CQG#1#2#3{Class. Quantum Grav. {\xbold#1} ({\xold#2}) {\xold#3}}

\def\JHEP{\jhep}

\def\MPLA#1#2#3{Mod. Phys. Lett. {\xbf A}{\xbold#1} ({\xold#2}) {\xold#3}}

\def\PLB#1#2#3{Phys. Lett. {\xbf B}{\xbold#1} ({\xold#2}) {\xold#3}}

\def\PRD#1#2#3{Phys. Rev. {\xbf D}{\xbold#1} ({\xold#2}) {\xold#3}}

\newcount\sectioncount
\sectioncount=0
\def\section#1#2{\global\eqcount=0
	\global\subsectioncount=0
        \global\advance\sectioncount by 1
	\ifnum\sectioncount>1
	        \vskip2\baselineskip
	\fi
\noindent{\twelvecp\the\sectioncount. #2}\par\nobreak
       \vskip.5\baselineskip\noindent
        \xdef#1{{\old\the\sectioncount}}}
\newcount\subsectioncount
\def\subsection#1#2{\global\advance\subsectioncount by 1
\vskip.75\baselineskip\noindent\line{\tencp\the\sectioncount.\the\subsectioncount. #2\hfill}\nobreak\vskip.4\baselineskip\nobreak\noindent\xdef#1{{\old\the\sectioncount}.{\old\the\subsectioncount}}}
\def\immediatesubsection#1#2{\global\advance\subsectioncount by 1
\vskip-\baselineskip\noindent
\line{\tencp\the\sectioncount.\the\subsectioncount. #2\hfill}
	\vskip.5\baselineskip\noindent
	\xdef#1{{\old\the\sectioncount}.{\old\the\subsectioncount}}}
\newcount\subsubsectioncount
\def\subsubsection#1#2{\global\advance\subsubsectioncount by 1
\vskip.75\baselineskip\noindent\line{\tencp\the\sectioncount.\the\subsectioncount.\the\subsubsectioncount. #2\hfill}\nobreak\vskip.4\baselineskip\nobreak\noindent\xdef#1{{\old\the\sectioncount}.{\old\the\subsectioncount}.{\old\the\subsubsectioncount}}}
\newcount\appendixcount
\appendixcount=0
\def\appendix#1{\global\eqcount=0
        \global\advance\appendixcount by 1
        \vskip2\baselineskip\noindent
        \ifnum\the\appendixcount=1
        {\twelvecp Appendix A: #1}\par\nobreak
                        \vskip.5\baselineskip\noindent\fi
        \ifnum\the\appendixcount=2
        {\twelvecp Appendix B: #1}\par\nobreak
                        \vskip.5\baselineskip\noindent\fi
        \ifnum\the\appendixcount=3
        {\twelvecp Appendix C: #1}\par\nobreak
                        \vskip.5\baselineskip\noindent\fi}
\def\acknowledgements{\immediate\write\contentswrite{\item{}\hbox
        to\contentlength{Acknowledgements\dotfill\the\pageno}}
        \vskip2\baselineskip\noindent
        \underbar{\it Acknowledgements:}\ }
\newcount\eqcount
\eqcount=0
\def\Eqn#1{\global\advance\eqcount by 1
\ifnum\the\sectioncount=0
	\xdef#1{{\noexpand\oldsize\the\eqcount}}
	\eqno({\oldstyle\the\eqcount})
\else
        \ifnum\the\appendixcount=0
\xdef#1{{\noexpand\oldsize\the\sectioncount}.{\noexpand\oldsize\the\eqcount}}
                \eqno({\oldstyle\the\sectioncount}.{\oldstyle\the\eqcount})\fi
        \ifnum\the\appendixcount=1
	        \xdef#1{{\noexpand\oldstyle A}.{\noexpand\oldstyle\the\eqcount}}
                \eqno({\oldstyle A}.{\oldstyle\the\eqcount})\fi
        \ifnum\the\appendixcount=2
	        \xdef#1{{\noexpand\oldstyle B}.{\noexpand\oldstyle\the\eqcount}}
                \eqno({\oldstyle B}.{\oldstyle\the\eqcount})\fi
        \ifnum\the\appendixcount=3
	        \xdef#1{{\noexpand\oldstyle C}.{\noexpand\oldstyle\the\eqcount}}
                \eqno({\oldstyle C}.{\oldstyle\the\eqcount})\fi
\fi}
\def\eqn{\global\advance\eqcount by 1
\ifnum\the\sectioncount=0
	\eqno({\oldstyle\the\eqcount})
\else
        \ifnum\the\appendixcount=0
                \eqno({\oldstyle\the\sectioncount}.{\oldstyle\the\eqcount})\fi
        \ifnum\the\appendixcount=1
                \eqno({\oldstyle A}.{\oldstyle\the\eqcount})\fi
        \ifnum\the\appendixcount=2
                \eqno({\oldstyle B}.{\oldstyle\the\eqcount})\fi
        \ifnum\the\appendixcount=3
                \eqno({\oldstyle C}.{\oldstyle\the\eqcount})\fi
\fi}
\def\multi{\global\advance\eqcount by 1}
\def\multieqn#1{({\oldstyle\the\sectioncount}.{\oldstyle\the\eqcount}\hbox{#1})}
\def\multiEqn#1#2{\xdef#1{{\oldstyle\the\sectioncount}.{\old\the\eqcount}#2}
        ({\oldstyle\the\sectioncount}.{\oldstyle\the\eqcount}\hbox{#2})}
\def\multiEqnAll#1{\xdef#1{{\oldstyle\the\sectioncount}.{\old\the\eqcount}}}
\newcount\tablecount
\tablecount=0
\def\Table#1#2#3{\global\advance\tablecount by 1
\immediate\write\intrefwrite{\def\noexpand#1{{\noexpand\oldsize\the\tablecount}}}
       \vtop{\vskip2\parskip
       \centerline{#2}
       \vskip5\parskip
       \centerline{\it Table \the\tablecount: #3}
       \vskip2\parskip}}
\newcount\figurecount
\figurecount=0
\def\Figure#1#2#3{\global\advance\figurecount by 1
\immediate\write\intrefwrite{\def\noexpand#1{{\noexpand\oldsize\the\figurecount}}}
       \vtop{\vskip2\parskip
       \centerline{#2}
       \vskip4\parskip
       \centerline{\it Figure \the\figurecount: #3}
       \vskip3\parskip}}
\newtoks\url
\def\Href#1#2{\catcode`\#=12\url={#1}\catcode`\#=\active#2}
\def\href#1#2{{#2}}

\parskip=3.5pt plus .3pt minus .3pt
\baselineskip=14pt plus .1pt minus .05pt
\lineskip=.5pt plus .05pt minus .05pt
\lineskiplimit=.5pt
\abovedisplayskip=18pt plus 4pt minus 2pt
\belowdisplayskip=\abovedisplayskip
\hsize=14cm
\vsize=19cm
\hoffset=1.5cm
\voffset=1.8cm
\frenchspacing
\footline={}
\raggedbottom

\newskip\origparindent
\origparindent=\parindent

\def\ss{\scriptstyle}

\def\*{\partial}
\def\punkt{\,\,.}
\def\komma{\,\,,}

\def\={\!=\!}
\def\small#1{{\hbox{$#1$}}}

\def\fraction#1{\small{1\over#1}}
\def\fr{\fraction}
\def\Fraction#1#2{\small{#1\over#2}}
\def\Fr{\Fraction}

\def\eg{{\it e.g.}}

\def\ie{{\it i.e.}}

\def\a{\alpha}

\def\l{\lambda}

\def\lra{\longrightarrow}
\def\lla{\longleftarrow}

\def\rarrowover#1{\vtop{\baselineskip=0pt\lineskip=0pt
      \ialign{\hfill##\hfill\cr$\ra$\cr$#1$\cr}}}

\def\larrowover#1{\vtop{\baselineskip=0pt\lineskip=0pt
      \ialign{\hfill##\hfill\cr$\la$\cr$#1$\cr}}}

\def\CC{{\Bbb C}}
\def\HH{{\Bbb H}}


\def\appendix#1#2{\global\eqcount=0
        \global\advance\appendixcount by 1
        \vskip2\baselineskip\noindent
        \ifnum\the\appendixcount=1
        \immediate\write\intrefwrite{\def\noexpand#1{A}}
        {\twelvecp Appendix A: #2}\par\nobreak
                        \vskip.5\baselineskip\noindent\fi
        \ifnum\the\appendixcount=2
        {\twelvecp Appendix B: #2}\par\nobreak
                        \vskip.5\baselineskip\noindent\fi
        \ifnum\the\appendixcount=3
        {\twelvecp Appendix C: #2}\par\nobreak
                        \vskip.5\baselineskip\noindent\fi}

\def\lb{\bar\l}

\def\textfrac#1#2{\raise .45ex\hbox{\the\scriptfont0 #1}\nobreak\hskip-1pt/\hskip-1pt\hbox{\the\scriptfont0 #2}}


\def\frac{\Fr}

\def\mathbb{\Bbb}






\catcode`@=11
\def\openupnormal{\afterassignment\@penupnormal\dimen@=}
\def\@penupnormal{\advance\normallineskip\dimen@
  \advance\normalbaselineskip\dimen@
  \advance\normallineskiplimit\dimen@}
\catcode`@=12

\def\EqMatrix{\let\quad\enspace\openupnormal6pt\matrix}

%
\line{
\epsfysize=18mm
\epsffile{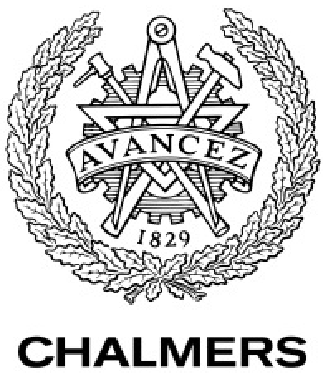}
\hfill}
\vskip-16mm

\line{\hfill}
\line{\hfill Gothenburg preprint}
\line{\hfill December, {\old2017}}
\line{\hrulefill}


\headtext={Cederwall: 
``Pure spinor superspace action for {\eightit D}=6, {\eightit N}=1 SYM''}

\vfill

\centerline{\sixteenhelvbold
Pure spinor superspace action}

\vskip3\parskip

\centerline{\sixteenhelvbold
for D=6, N=1 super-Yang--Mills theory}

%

\vfill

\centerline{\twelvehelvbold Martin Cederwall}

\vfill
\vskip-1cm

\centerline{\it Division for Theoretical Physics}
\centerline{\it Department of Physics}
\centerline{\it Chalmers University of Technology}
\centerline{\it SE 412 96 Gothenburg, Sweden}

\vfill

{\narrower\noindent \underbar{Abstract:}
A Batalin--Vilkovisky action for $D=6$, $N=1$ super-Yang--Mills theory,
including coupling to hypermultiplets, is given. The formalism
involves pure spinor superfields.
The geometric properties of the $D=6$, $N=1$ pure spinors (which
differ from Cartan
pure spinors) are examined.
Unlike the situation for maximally supersymmetric models, the fields
and antifields (including ghosts) of the vector
multiplet reside in separate superfields.
The formalism
provides an off-shell superspace formulation for matter
hypermultiplets, which in a traditional treatment are on-shell.

\smallskip}
\vfill

\font\xxtt=cmtt6

\vtop{\baselineskip=.6\baselineskip\xxtt
\line{\hrulefill}
\catcode`\@=11
\line{email: martin.cederwall@chalmers.se\hfill}
\catcode`\@=\active
}

\eject



\def\lb{\bar\l}

\def\textfrac#1#2{\raise .45ex\hbox{\the\scriptfont0 #1}\nobreak\hskip-1pt/\hskip-1pt\hbox{\the\scriptfont0 #2}}


\def\frac{\Fr}

\def\mathbb{\Bbb}

\def\larrowover#1{\vtop{\baselineskip=0pt\lineskip=0pt
      \ialign{\hfill##\hfill\cr$\lla$\cr$#1$\cr}}}
\def\rarrowover#1{\vtop{\baselineskip=0pt\lineskip=0pt
      \ialign{\hfill##\hfill\cr$\lra$\cr$#1$\cr}}}

\newskip\scrskip
\scrskip=-.6pt plus 0pt minus .1pt


\newwrite\intrefwrite
\immediate\openout\intrefwrite=\jobname.intref

\newwrite\contentswrite

\newdimen\sublength
\sublength=\hsize 
\advance\sublength by -\parindent

\newdimen\contentlength
\contentlength=\sublength

\advance\sublength by -\parindent

\def\section#1#2{\global\eqcount=0
	\global\subsectioncount=0
        \global\advance\sectioncount by 1
\ifnum\the\sectioncount=1\immediate\openout\contentswrite=\jobname.contents\fi
\immediate\write\contentswrite{\item{\the\sectioncount.}\hbox to\contentlength{#2\dotfill\the\pageno}}
	\ifnum\sectioncount>1
	        \vskip2\baselineskip
	\fi
\immediate\write\intrefwrite{\def\noexpand#1{{\noexpand\oldsize\the\sectioncount}}}\noindent{\twelvecp\the\sectioncount. #2}\par\nobreak
       \vskip.5\baselineskip\noindent}

\def\subsection#1#2{\global\advance\subsectioncount by 1
\immediate\write\contentswrite{\itemitem{\the\sectioncount.\the\subsectioncount.}\hbox
to\sublength{#2\dotfill\the\pageno}}
\immediate\write\intrefwrite{\def\noexpand#1{{\noexpand\oldsize\the\sectioncount}.{\noexpand\oldsize\the\subsectioncount}}}\vskip.75\baselineskip\noindent\line{\tencp\the\sectioncount.\the\subsectioncount. #2\hfill}\nobreak\vskip.4\baselineskip\nobreak\noindent}

\def\immediatesubsection#1#2{\global\advance\subsectioncount by 1
\immediate\write\contentswrite{\itemitem{\the\sectioncount.\the\subsectioncount.}\hbox
to\sublength{#2\dotfill\the\pageno}}
\immediate\write\intrefwrite{\def\noexpand#1{{\noexpand\oldsize\the\sectioncount}.{\noexpand\oldsize\the\subsectioncount}}}
\vskip-\baselineskip\noindent
\line{\tencp\the\sectioncount.\the\subsectioncount. #2\hfill}
	\vskip.5\baselineskip\noindent}

\def\contentsout{\catcode`\@=11
        \vskip2\baselineskip
        {\noindent\twelvecp Contents}\hfill\vskip\baselineskip
        \input\jobname.contents
        \catcode`\@=\active\rm
\vskip3\baselineskip
}

\def\refout{\eightpoint\catcode`\@=11
        \immediate\write\contentswrite{\item{}\hbox to\contentlength{References\dotfill\the\pageno}}
        \xrm\immediate\closeout\refwrite
        \vskip2\baselineskip
        {\noindent\twelvecp References}\hfill\vskip\baselineskip
        \baselineskip=.75\baselineskip
        \input\jobname.refs
        \baselineskip=4\baselineskip \divide\baselineskip by 3
        \catcode`\@=\active\rm}

\def\cf{{\it cf}}


\ref\BerkovitsI{N. Berkovits, 
{\xit ``Super-Poincar\'e covariant quantization of the superstring''}, 
\jhep{00}{04}{2000}{018} [\hepth{0001035}].}

\ref\BerkovitsParticle{N. Berkovits, {\xit ``Covariant quantization of
the superparticle using pure spinors''}, \jhep{01}{09}{2001}{016}
[\hepth{0105050}].}

\ref\CederwallNilssonTsimpisI{M. Cederwall, B.E.W. Nilsson and D. Tsimpis,
{\xit ``The structure of maximally supersymmetric super-Yang--Mills
theory --- constraining higher order corrections''},
\jhep{01}{06}{2001}{034} 
[\hepth{0102009}].}

\ref\CederwallNilssonTsimpisII{M. Cederwall, B.E.W. Nilsson and D. Tsimpis,
{\xit ``D=10 super-Yang--Mills at $\ss O(\a'^2)$''},
\JHEP{01}{07}{2001}{042} [\hepth{0104236}].}

\ref\SpinorialCohomology{M. Cederwall, B.E.W. Nilsson and D. Tsimpis, 
{\xit ``Spinorial cohomology and maximally supersymmetric theories''},
\jhep{02}{02}{2002}{009} [\hepth{0110069}];
M. Cederwall, {\xit ``Superspace methods in string theory,
supergravity and gauge theory''}, Lectures at the XXXVII Winter School
in Theoretical Physics ``New Developments in Fundamental Interactions
Theories'',  Karpacz, Poland,  Feb. 6-15, 2001, \hepth{0105176}.}

\ref\CederwallBLG{M. Cederwall, {\xit ``N=8 superfield formulation of
the Bagger--Lambert--Gustavsson model''}, \jhep{08}{09}{2008}{116}
[\arxiv{0808}{3242}].}

\ref\CederwallABJM{M. Cederwall, {\xit ``Superfield actions for N=8 
and N=6 conformal theories in three dimensions''},
\jhep{08}{10}{2008}{70}
[\arxiv{0809}{0318}].}

\ref\PureSGI{M. Cederwall, {\xit ``Towards a manifestly supersymmetric
    action for D=11 supergravity''}, \jhep{10}{01}{2010}{117}
    [\arxiv{0912}{1814}].}  

\ref\PureSGII{M. Cederwall, 
{\xit ``D=11 supergravity with manifest supersymmetry''},
    \MPLA{25}{2010}{3201} [\arxiv{1001}{0112}].}

\ref\PureSpinorOverview{M. Cederwall, {\xit ``Pure spinor superfields
--- an overview''}, Springer Proc. Phys. {\xbf153} ({\xrm2013}) {\xrm61} 
[\arxiv{1307}{1762}].}

\ref\KacSuperalgebras{V.G. Kac, {\xit ``Classification of simple Lie
superalgebras''}, Funktsional. Anal. i Prilozhen. {\xbold9}
({\xold1975}) {\xold91}.}

\ref\CederwallExceptionalTwistors{M. Cederwall, {\xit ``Twistors and
supertwistors for exceptional field
theory''}, \jhep{15}{12}{2015}{123} [\arxiv{1510}{02298}].}

\ref\MaDBrane{C.-T. Ma, {\xit ``Gauge transformation of double field
theory for open string''}, \PRD{92}{2015}{066004} [\arxiv{1411}{0287}].}

\ref\BandosStringsSuperspace{I Bandos, {\xit ``Strings in doubled
superspace''}, \PLB{751}{2015}{402} [\arxiv{1507}{07779}].}

\ref\BandosESeven{I Bandos, {\xit ``On section conditions of
$E_{7(+7)}$ exceptional field theory and superparticle in N=8 central
charge superspace''}, \jhep{16}{01}{2016}{132} [\arxiv{1512}{02287}].}


\ref\CederwallNilssonSix{M. Cederwall and B.E.W. Nilsson, {\xit ``Pure
spinors and D=6 super-Yang--Mills''}, \arxiv{0801}{1428}.}

\ref\CGNN{M. Cederwall, U. Gran, M. Nielsen and B.E.W. Nilsson, 
{\xit ``Manifestly supersymmetric M-theory''}, 
\JHEP{00}{10}{2000}{041} [\hepth{0007035}];
{\xit ``Generalised 11-dimensional supergravity''}, \hepth{0010042}.}

\ref\CederwallNilssonTsimpisI{M. Cederwall, B.E.W. Nilsson and D. Tsimpis,
{\xit ``The structure of maximally supersymmetric super-Yang--Mills
theory --- constraining higher order corrections''},
\jhep{01}{06}{2001}{034} 
[\hepth{0102009}].}

\ref\CederwallNilssonTsimpisII{M. Cederwall, B.E.W. Nilsson and D. Tsimpis,
{\xit ``D=10 super-Yang--Mills at $\ss O(\a'^2)$''},
\JHEP{01}{07}{2001}{042} [\hepth{0104236}].}

\ref\SpinorialCohomology{M. Cederwall, B.E.W. Nilsson and D. Tsimpis, 
{\xit ``Spinorial cohomology and maximally supersymmetric theories''},
\jhep{02}{02}{2002}{009} [\hepth{0110069}].}

\ref\SuperspaceMethods{M. Cederwall,
{\xit ``Superspace methods in string theory,
supergravity and gauge theory''}, Lectures at the XXXVII Winter School
in Theoretical Physics ``New Developments in Fundamental Interactions
Theories'',  Karpacz, Poland,  Feb. 6-15, 2001, \hepth{0105176}.}

\ref\CederwallNilssonTsimpisIII{M. Cederwall, B.E.W. Nilsson and D. Tsimpis,
{\xit ``Spinorial cohomology of abelian D=10 super-Yang--Mills at $\ss
O(\a'^3)$''}, 
\JHEP{02}{11}{2002}{023} [\hepth{0205165}].}

\ref\CGNT{M. Cederwall, U. Gran, B.E.W. Nilsson and D. Tsimpis,
{\xit ``Supersymmetric corrections to eleven-dimen\-sional supergravity''},
\jhep{05}{05}{2005}{052} [\hepth{0409107}].}

\ref\CederwallBLG{M. Cederwall, {\xit ``N=8 superfield formulation of
the Bagger--Lambert--Gustavsson model''}, \jhep{08}{09}{2008}{116}
[\arxiv{0808}{3242}].}

\ref\CederwallABJM{M. Cederwall, {\xit ``Superfield actions for N=8 
and N=6 conformal theories in three dimensions''},
\jhep{08}{10}{2008}{70}
[\arxiv{0809}{0318}].}

\ref\CederwallThreeConf{M. Cederwall, {\xit ``Pure spinor superfields,
with application to D=3 conformal models''}, 
Proc. Estonian Acad. Sci. {\xbf4} ({\xold2010})
[\arxiv{0906}{5490}].}

\ref\PureSGI{M. Cederwall, {\xit ``Towards a manifestly supersymmetric
    action for D=11 supergravity''}, \jhep{10}{01}{2010}{117}
    [\arxiv{0912}{1814}].}  

\ref\PureSGII{M. Cederwall, 
{\xit ``D=11 supergravity with manifest supersymmetry''},
    \MPLA{25}{2010}{3201} [\arxiv{1001}{0112}].}

\ref\SupergeometryPureSpinors{M. Cederwall, {\xit ``From supergeometry
to pure spinors''}, in the proceedings of the 6th Mathematical physics
meeting, Belgrade, September 2010, \arxiv{1012}{3334}.}

\ref\CederwallKarlssonBI{M. Cederwall and A. Karlsson, {\xit ``Pure
spinor superfields and Born--Infeld theory''},
\jhep{11}{11}{2011}{134} [\arxiv{1109}{0809}].}

\ref\CederwallPureSpinorSpace{M. Cederwall, {\xit ``The geometry of
pure spinor space''}, \jhep{12}{01}{2012}{150}  
\hfill\break[\arxiv{1111}{1932}].}

\ref\CederwallKarlssonLoop{M. Cederwall and A. Karlsson, {\xit ``Loop
amplitudes in maximal supergravity with manifest supersymmetry''},
\jhep{13}{03}{2013}{114} [\arxiv{1212}{5175}].}

\ref\PureSpinorOverview{M. Cederwall, {\xit ``Pure spinor superfields
--- an overview''}, Springer Proc. Phys. {\xbf153} ({\xrm2013}) {\xrm61} 
[\arxiv{1307}{1762}].}

\ref\ChangDeformationsI{C.-M. Chang, Y.-H. Lin, Y. Wang and X. Yin,
{\xit ``Deformations with maximal supersymmetries part 1: on-shell
formulation''},  \arxiv{1403}{0545}.}

\ref\ChangDeformationsII{C.-M. Chang, Y.-H. Lin, Y. Wang and X. Yin,
{\xit ``Deformations with maximal supersymmetries part 2: off-shell
formulation''}, \jhep{16}{04}{2016}{171} [\arxiv{1403}{0709}].}

\ref\CederwallReformulation{M. Cederwall, {\xit ``An off-shell superspace
reformulation of $D=4$, $N=4$ super-Yang--Mills theory''}, \arxiv{1707}{00554}.}

\ref\GIKOS{A. Galperin, E. Ivanov, S. Kalitzin, V. Ogievetsky and
E. Sokatchev, {\xit ``Unconstrained $N=2$ matter,
Yang--Mills and supergravity theories in harmonic
superspace''}, \CQG1{1984}{469}.} 

\ref\ChicherinSokatchevI{D. Chicherin and E. Sokatchev, {\xit ``$N=4$
super-Yang--Mills in LHC superspace. Part I: Classical and quantum
theory''}, \jhep{17}{02}{2017}{062} [\arxiv{1601}{06803}].}

\ref\ChicherinSokatchevII{D. Chicherin and E. Sokatchev, {\xit ``$N=4$
super-Yang--Mills in LHC superspace. Part II: Non-chiral correlation functions
of the stress-tensor multiplet''}, \jhep{17}{03}{2017}{048}
[\arxiv{1601}{06804}].} 

\ref\BerkovitsGreenRussoVanhove{N. Berkovits, M.B. Green, J. Russo and
    P. Vanhove, {\xit ``Non-renormalization conditions for four-gluon
    scattering in supersymmetric string and field
    theory''}, \jhep{09}{11}{2009}{063} [\arxiv{0908}{1923}].}

\ref\BerkovitsNonMinimal{N. Berkovits,
{\xit ``Pure spinor formalism as an N=2 topological string''},
\jhep{05}{10}{2005}{089} [\hepth{0509120}].}

\ref\BerkovitsNekrasovMultiloop{N. Berkovits and N. Nekrasov, {\xit
    ``Multiloop superstring amplitudes from non-minimal pure spinor
    formalism''}, \jhep{06}{12}{2006}{029} [\hepth{0609012}].}

\ref\ChangDeformationsII{C.-M. Chang, Y.-H. Lin, Y. Wang and X. Yin,
{\xit ``Deformations with maximal supersymmetries part 2: off-shell
formulation''}, \jhep{16}{04}{2016}{171} [\arxiv{1403}{0709}].}

\ref\CederwallPalmkvistBorcherds{M. Cederwall and J. Palmkvist, {\xit
``Superalgebras, constraints and partition functions''},
\jhep{08}{15}{2015}{36} [\arxiv{1503}{06215}].}

\ref\BermanCederwallKleinschmidtThompson{D.S. Berman, M. Cederwall,
A. Kleinschmidt and D.C. Thompson, {\xit ``The gauge structure of
generalised diffeomorphisms''}, \jhep{13}{01}{2013}{64} [\arxiv{1208}{5884}].}

\ref\BuchbinderIvanovMerzlikin{I.L. Buchbinder, E.A. Ivanov and
B.S. Merzlikin, {\xit ``Leading low-energy effective action in 6D
$N=1$ SYM theory with hypermultiplets''}, \arxiv{1711}{03302}.}


\def\ZZZ{{\cal Z}}

\def\bl{\bar\lambda}

\contentsout

\section\IntroductionSection{Introduction}Pure spinor superfields
(see ref. [\PureSpinorOverview] and references therein)
have
been used in the construction of actions for maximally supersymmetric
theories
[\CederwallBLG\skipref\CederwallABJM\skipref\CederwallThreeConf\skipref\PureSGI\skipref\PureSGII\skipref\SupergeometryPureSpinors\skipref\CederwallKarlssonBI\skipref\CederwallPureSpinorSpace\skipref\CederwallKarlssonLoop\skipref\ChangDeformationsI\skipref\ChangDeformationsII--\CederwallReformulation].
It is there that the formalism, originating in superstring
theory [\BerkovitsI\skipref\BerkovitsParticle\skipref\BerkovitsNonMinimal--\BerkovitsNekrasovMultiloop] and in the deformation theory for maximally supersymmetric
super-Yang--Mills theory (SYM) and supergravity
[\CGNN\skipref\CederwallNilssonTsimpisI\skipref\CederwallNilssonTsimpisII\skipref\SpinorialCohomology\skipref\SuperspaceMethods\skipref\CederwallNilssonTsimpisIII--\CGNT],
has its greatest power. The superspace
constraints, turned into a relation of the form ``$Q\Psi+\ldots=0$'',
where $Q$ is nilpotent, become the equations of motion in a
Batalin--Vilkovisky (BV) framework. Not only does this allow for a solution
to the long-standing problem of off-shell formulation of maximally
supersymmetric theories, the actions are also typically of a very
simple kind. Generically, they turn out to be of finite and low order
in the fields, even when the component field dynamics is
non-polynomial.

Surprisingly little work has been done on pure spinor superfields for
models with less than maximal supersymmetry. A classical description
of $D=6$, $N=1$ super-Yang--Mills theory was given in ref.
[\CederwallNilssonSix]. It was
based on minimal pure spinor variables, which precludes the treatment
of important issues like integration. It is the aim of the present
work to take this construction to the level of an action
principle. Such a formulation may, after gauge fixing, be used for
quantum calculations, to be compared \eg\ to the ones performed in
harmonic superspace [\BuchbinderIvanovMerzlikin].



\section\PureSpinorSection{$D=6$, $N=1$ pure spinors}

\immediatesubsection\MinimalSS{Minimal pure spinor variables}The
important
property for pure spinors in relation to supersymmetry
is the constraint
$$
(\lambda\gamma^a\lambda)=0\punkt\Eqn\PureSpinorConstraint
$$
When the anticommutator of two fermionic covariant derivatives
contains the torsion
$T_{\alpha\beta}{}^a=2\gamma^a_{\alpha\beta}$, this 
ensures that the BRST operator
$$
q=\lambda^\alpha D_\alpha\eqn
$$
is nilpotent, and (physical) fields may be defined as belonging to
some cohomology of $q$. The pure spinor $\lambda$ carries ghost number one.

The $D=6$, $N=1$ spinors transform under $Spin(1,5)\times SU(2)$, the
latter being the R-symmetry group. For Minkowski signature, this
allows for (pseudo-)real 8-dimensional
chiral spinor representations in the form of
so called $SU(2)$-Majorana spinors. A convenient way to represent them
is as two-component spinors with quaternionic entries. One then uses
the isomorphism $SL(2;\HH)\approx Spin(1,5)$, and the R-symmetry
$SU(2)$ acts by quaternionic multiplication with elements of unit norm
from the right. We will
use this language only occasionally, but instead work with matrices
$(\gamma^a)_{\alpha\beta}$ or $(\gamma^a)^{\alpha\beta}$,
$a=1,\ldots,6$,
acting on the
respective chiral spinors, and $(\sigma_i)^\alpha{}_\beta$ or
$(\sigma_i)_\alpha{}^\beta$, $i=1,2,3$. In the quaternionic language,
the latter are identified with right multiplication by $-e_i$, the
imaginary quaternionic units. They satisfy
$\sigma_i\sigma_j=-\delta_{ij}+\epsilon_{ijk}\sigma_k$.
Some more spinor identities are collected in
Appendix \GammaMatrixAppendix.
The numbering for Dynkin labels is that of Figure \DynkinFigure, where an
upper spinor index is represented by $(001)(1)$.

\Figure\DynkinFigure{\epsffile{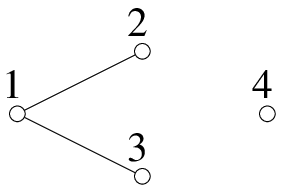}}{Labelling of the Dynkin diagram for
$D_3\times A_1$.}

The symmetry properties of spinor bilinears are:
$$
\matrix{
\hfill\hbox{symmetric:}\quad&\quad(\gamma^a)_{\alpha\beta}\quad\hfill
                      &\quad(\gamma^{abc}\sigma_i)_{\alpha\beta}\hfill\cr
\hfill\hbox{antisymmetric:}\quad&\quad(\gamma^a\sigma_i)_{\alpha\beta}
                        \quad\hfill
                      &\quad(\gamma^{abc})_{\alpha\beta}\hfill\cr
}\eqn
$$

A bosonic spinor $\lambda^\alpha$ in $({\bf4},{\bf2})=(001)(1)$, subject to
the pure spinor constraint (\PureSpinorConstraint), will only yield
the single representation $(00n)(n)$ in its $n$'th power.
Counting the dimensions of these representations immediately gives the
partition function for the pure spinor (\cf. refs.
[\BerkovitsNekrasovMultiloop,\BermanCederwallKleinschmidtThompson,\CederwallPalmkvistBorcherds])
$$
Z(t)=\sum\limits_{n=0}^\infty(n+1){n+3\choose3}
={1+3t\over(1-t)^5}={1-6t^2+8t^3-3t^4\over(1-t)^8}\punkt
\Eqn\UnRefinedZ
$$
A refined partition function, counting the actual representation
content at each level, is given by
$$
\ZZZ(t)=\sum\limits_{n=0}^\infty(00n)(n)t^n
=\ZZZ_0(t)\otimes\left[
(000)(0)-(100)(0)t^2+(010)(1)t^3-(000)(2)t^4\right]\komma\Eqn\RefinedZ
$$
where $\ZZZ_0$ is the partition function for an unconstrained spinor,
$$
\ZZZ_0(t)=\sum\limits_{n=0}^\infty\otimes_s^n(001)(1)t^n\punkt\eqn
$$

As usual, the second factor in eq. (\RefinedZ) encodes the zero mode
cohomology of the BRST operator $q$, which will be described in
Section \DSixSYMSection.

An
attempt to solve the pure spinor constraint immediately shows that
complex pure spinors are needed. The manifold of pure spinors is a
5-dimensional complex manifold. The dimensionality is reflected in the
power of the denominator of eq. (\UnRefinedZ).
If one considers a complex spinor as a bifundamental $\lambda^{Aa}$
of $SU(4)\times SU(2)$, the pure spinor constraint takes the form
$\epsilon_{ab}\lambda^{Aa}\lambda^{Bb}=0$. Obviously, any spinor of
the form
$\lambda=\left(\matrix{\ell^A,0}\right)$ is pure\foot{This amounts to
the statement that any $Spin(6)$ spinor is pure, in the sense of
Cartan.}, and all solutions
can be obtained from this solution by transformations in $SU(4)\times
SU(2)$. This tells us that the space of pure spinors is
$\CC^4\times\CC P^1$.

The conjugate variable to $\lambda^\alpha$,
$\omega_\alpha={\*\over\*\lambda^\alpha}$ is not well defined, since
it does not preserve the pure spinor constraint. However, the
operators
$$
N=(\lambda\omega)\komma\quad
N^{ab}=(\lambda\gamma^{ab}\omega)\komma\quad
N^i=(\lambda\sigma^i\omega)\eqn
$$
are well defined.

\subsection\NonMinimalSS{Non-minimal variables and integration}For
several reasons, it is necessary to include non-minimal variables
[\BerkovitsNonMinimal], a
bosonic variable $\bl_\alpha$ and the ``fermionic''
$r_\alpha=d\bl_\alpha$. One reason is the construction of a
non-degenerate integration measure, another, as we will see, is the
need for operators with negative ghost number.
The BRST operator is modified to
$$
Q=q+\bar\*=\lambda^\alpha D_\alpha+d\bl_\alpha{\*\over\*\bl_\alpha}\punkt\eqn
$$
$\bl$ can be considered as the complex conjugate of $\lambda$. It is
pure, and differentiation gives $(\bl\gamma^ad\bl)=0$.

If superfields are functions of the non-minimal variables $x^a$,
$\theta^\alpha$, $\lambda^\alpha$, $\bl_\alpha$ and $d\bl_\alpha$,
they are forms with antiholomorphic indices on complex pure spinor
space.
A tentative integration can then be taken as
$$
\int[dZ]\phi\sim\int d^6x\,d^8\theta\int\Omega\wedge\phi\komma\eqn
$$
if it is possible to find a holomorphic 5-form $\Omega$.

From the description of pure spinor space as $\CC^4\times\CC P^1$, it
is clear that there is not only one, but three holomorphic 5-forms,
which can be written as $d^4y\,z^pdz$, $p=0,1,2$, where $y$ parametrises
$\CC^4$ and $z$ $\CC P^1$. They transform as a triplet under
R-symmetry.
We will in fact use the full triplet, and have a ``triplet
integration''. It will become clear, when actions are formed in
Section \BVActionSection,
that this is necessary in order to
maintain covariance, and absorb transformations of diverse fields.

For our purposes, and a closer correspondence with the cohomology of
Section \DSixSYMSection, we will write down an expression for the
holomorphic 5-forms $\Omega_i$ in a fully covariant way. They are
$$
\Omega_i=(\lambda\bl)^{-1}(\bl\sigma^jd\lambda)
(d\lambda\gamma^a\sigma_jd\lambda)(d\lambda\gamma_a\sigma_id\lambda)
\punkt\eqn
$$
Although $\bl$ is used to form a covariant expression, it can be
checked that $\bar\partial\Omega=0$. In addition, the forms satisfy
$$
(\sigma^i\lambda)^\alpha\Omega_i=0\punkt\Eqn\LambdaOmega
$$
Except for the presence of a triple of holomorphic top-forms instead
of single one, this mirrors closely the construction for $D=10$ pure
spinor superfields. 
As we will see in the following section, the integration measure is
directly connected to the highest cohomology of a scalar pure spinor
superfield, which is the present case will be the triplet of auxiliary
fields $H_i$ in the super-Yang--Mills multiplet.

The geometry corresponding to the integration at hand, with a volume
form $\hbox{Vol}=\Omega_i\wedge\bar\Omega_i$, is not the one inherited by
embedding pure spinor space in flat spinor space. The latter one would
scale like $d\lambda^5d\bl^5$, while the actual volume scales like
$\hbox{Vol}\sim\lambda^{-1}\bl^{-1}d\lambda^5d\bl^5$. This is quite
similar to the 10-dimensional situation [\CederwallPureSpinorSpace].
As usual, integrals have to be regularised by a factor
$\hbox{exp}\{Q,\chi\}$. A convenient choice is $\chi=-(\lb\theta)$, giving
$$
e^{\{Q,\chi\}}=e^{-(\lambda\lb)-(d\lb\theta)}\komma
\Eqn\Regulator
$$
which both regulates the
integral over pure spinor space at infinity and saturates the
fermionic integral.

\section\DSixSYMSection{Cohomology and supermultiplets}In this section
we will construct pure spinor superfields containing the off-shell
SYM multiplet and its current multiplet, and the on-shell hypermultiplet.


\subsection\VectorSS{The vector multiplet}The standard superspace
treatment of supersymmetric gauge theory formulates SYM as gauge
theory on superspace. A connection 1-form $A$ is decomposed as
$A=E^aA_a(x,\theta)+E^\alpha A_\alpha(x,\theta)$. The dimension 1 part of the
field strength $F_{\alpha\beta}$ is set to zero. This
contains two parts: a vector $(\gamma^a)^{\alpha\beta}F_{\alpha\beta}$
and a triplet of selfdual 3-forms
$(\gamma^{abc}\sigma_i)^{\alpha\beta}F_{\alpha\beta}$ in $(020)(2)$.
As usual, setting the vector to 0 is the conventional constraint,
expressing the superfield $A_a$, and thereby the entire field content,
in terms of the superfield $A_\alpha$.

One can now work with $A_\alpha$ alone. Consider a scalar pure spinor
superfield $\Psi(x,\theta,\lambda)$ of ghost number 1.
Its expansion in $\lambda$ contains the physical fields as
$\lambda^\alpha A_\alpha$. The (linearised) constraint on $F$ in
$(020)(2)$ now arises as the condition $q\Psi=0$. In addition, a
transformation $\delta\Psi=q\Lambda$ gives a gauge transformation, and
physical fields, modulo gauge transformations, arise as cohomology of
$q$.
It is well known that the relation $F_{\alpha\beta}=0$ does not imply
the field equations, but leaves the SYM fields off-shell.
Calculating the zero-mode cohomology indeed gives the SYM multiplet,
including the triplet $H_i$ of auxiliary fields, as shown in Table \SYMTable.

\Table\SYMTable{$$
\vtop{\baselineskip20pt\lineskip0pt
\ialign{
$\hfill#\quad$&$\,\hfill#\hfill\,$&$\,\hfill#\hfill\,$&$\,\hfill#\hfill\,$
&$\,\hfill#\hfill\,$&$\,\hfill#\hfill$\cr
            \hbox{gh\#}=&1    &0    &-1    &-2   \cr
\hbox{dim}=0&(000)(0)&&\phantom{(000)(0)}&\phantom{(000)(0)}\cr
        \fr2&\bullet&\bullet&               &       \cr
           1&\bullet&(100)(0)&\bullet&       &       \cr
       \Fr32&\bullet&(001)(1)&\bullet&\bullet      \cr
           2&\bullet&(000)(2)&\bullet&\bullet\cr
       \Fr52&\bullet&\bullet&\bullet&\bullet\cr
           3&\bullet&\bullet&\bullet&\bullet\cr
}}
$$}{The zero-mode cohomology of a scalar
superfield}

\noindent In this and the following tables, the representations and
quantum numbers (dimension, ghost number) of the
component fields are listed. 

Unlike the situation in $D=10$, where the SYM multiplet is an on-shell
multiplet, there is no cohomology at negative ghost numbers, which
also means that there is no room for differential constraints
(equations of motion) on the
physical fields. The equations of motion do not follow from
$Q\Psi=0$. Instead we will need some relation that effectively implies
the vanishing of the auxiliary fields. This will amount to finding an
operator $\hat H_i$ of ghost number $-1$ and dimension 2, the r\^ole of
which is to map the auxiliary fields to the ``beginning'' of the
superfield, and postulate $\hat H_i\Psi=0$. Such an operator will be
constructed in Section \BVActionSection.

\subsection\CurrentSS{The current (antifield) multiplet}The scalar
superfield of the previous subsection contains the ghost and the
physical off-shell SYM multiplet. In order to write a
Batalin--Vilkovisky action (Section \BVActionSection), also the
antifields for the fields and ghost are needed. They will come in a
field that is conjugate to $\Psi$ in the BV sense. This differs from
the situation in maximally supersymmetric SYM, where the scalar
superfield is self-conjugate, and $Q\Psi=0$ gives the equations of motion.

The antifield should have the auxiliary fields $H_i$ as its lowest
component, and must therefore itself be a triplet $\Psi^*_i$ with
ghost number $-1$ and dimension 2.
In order for a non-scalar superfield to carry a cohomology which is
not a product of its representation and the scalar cohomology, it has
to be subject to some condition. This has been encountered in a
number of situations [\CederwallBLG,\CederwallABJM,\PureSGI,\PureSGII,\CederwallKarlssonBI,\CederwallReformulation], and was named ``shift'' symmetry in
ref. [\CederwallKarlssonBI]. 
The appropriate condition is to consider the equivalence class
$$
\Psi^*_i\approx\Psi^*_i+(\lambda\sigma_i\zeta)\Eqn\AntiFieldShift
$$
for all possible spinor superfields $\zeta$. This will have
consequences for the cohomology. An immediate one is that the
zero-mode cohomology will contain $(\theta\sigma_i\chi^*)$, where
$\chi^*$ is the antifield for the physical spinor (acting with $Q$
gives precisely a shift as in eq. (\AntiFieldShift)). 
A complete calculation of the zero-mode cohomology yields Table
\SYMAntiTable, and the correct structure as the mirror of the fields
in table \SYMTable\ is reproduced.

\Table\SYMAntiTable{$$
\vtop{\baselineskip20pt\lineskip0pt
\ialign{
$\hfill#\quad$&$\,\hfill#\hfill\,$&$\,\hfill#\hfill\,$&$\,\hfill#\hfill\,$
&$\,\hfill#\hfill\,$&$\,\hfill#\hfill$\cr
        \hbox{gh\#}=    &-1    &-2    &-3    &-4   \cr
\hbox{dim}=2&(000)(2)&&\phantom{(000)(0)}&\phantom{(000)(0)}\cr
        \Fr52&(001)(1)&\bullet&               &       \cr
           3&(100)(1)&\bullet&\bullet&       &       \cr
       \Fr72&\bullet&\bullet&\bullet&\bullet      \cr
           4&\bullet&(000)(0)&\bullet&\bullet\cr
       \Fr92&\bullet&\bullet&\bullet&\bullet\cr
           5&\bullet&\bullet&\bullet&\bullet\cr
}}
$$}{The zero-mode cohomology of the triplet 
antifield}

It now becomes clear that the operator $\hat H_i$, needed to put the vector
multiplet on shell, should be an operator that maps the scalar field
$\Psi$ to a triplet field of the type described in the present
subsection.

We also note that the shift symmetry can be implemented in some
action, if the triplet integration and the antifield are used
together; an expression $\int[dZ]_i\Psi^*_i\ldots$ will automatically imply
it, since, as noted in Section \NonMinimalSS\ (eq. (\LambdaOmega)),
$[dZ]_i(\sigma_i\lambda)^\alpha\ldots=0$.

\subsection\HyperSS{The hypermultiplet}Finally, we give the superfield
corresponding to the hypermultiplet. There are no ghosts, so the
superfield should have as its lowest component the scalars of
dimension 1 and ghost number 0. The four scalars transform as $(2,2)$
under $SU(2)_L\times SU(2)_R$, where the second factor is an additional
$SU(2)$ R-symmetry leaving the vector multiplet inert.
It is convenient to collect them in a quaternion $\phi$, where the
``old'' $SU(2)_L$ acts by left multiplication and the new one by right
multiplication by unit quaternions.
We thus introduce a superfield $\Phi\in\HH$ with dimension 1 and ghost
number 0. It enjoys a shift symmetry
$$
\Phi\approx\Phi+\lambda^\dagger\rho\komma\eqn
$$
where now $\lambda$ is written in the quaternionic 2-component
notation described in Section \MinimalSS. The parameter
$\rho$ in the shift
term is a spinor transforming under the new R-symmetry from the right,
but inert under the old one. It implies the occurrence of such a
spinor in the zero-mode cohomology. The complete zero-mode cohomology
is displayed in Table \HyperTable.

\Table\HyperTable{$$
\vtop{\baselineskip20pt\lineskip0pt
\ialign{
$\hfill#\quad$&$\,\hfill#\hfill\,$&$\,\hfill#\hfill\,$&$\,\hfill#\hfill\,$
&$\,\hfill#\hfill\,$&$\,\hfill#\hfill$\cr
            \hbox{gh\#}=&0    &-1    &-2    &-3   \cr
\hbox{dim}=1&(000)(1)(1)&&\phantom{(000)(0)(0)}&\phantom{(000)(0)(0)}\cr
        \Fr32&(001)(0)(1)&\bullet&               &       \cr
           2&\bullet&\bullet&\bullet&       &       \cr
       \Fr52&\bullet&(010)(0)(1)&\bullet&\bullet      \cr
           3&\bullet&(000)(1)(1)&\bullet&\bullet\cr
       \Fr72&\bullet&\bullet&\bullet&\bullet\cr
           4&\bullet&\bullet&\bullet&\bullet\cr
}}
$$}{The zero-mode cohomology of the
hypermultiplet field}

The field is self-conjugate, in that it contains both the fields of the
hypermultiplet and their antifields, and the multiplet is of course an
on-shell multiplet in the traditional sense, and $Q\Phi=0$ implies the
component equations of motion.

\section\BVActionSection{Batalin--Vilkovisky actions}With the
description of the fields from Section \DSixSYMSection, we are now
ready to write down BV actions. We will begin with the linearised
theory, and then give the full interacting theory in
Section \InteractionSS.
A necessary ingredient will be certain operators, which are first
given in Section \OperatorSS.

The BV action will of course be a scalar. The consistency condition is
the BV master equation $(S,S)=0$. Some care has to be taken to define
the antibracket $(\cdot,\cdot)$, especially since the ``Lagrangian''
carries an $SU(2)$ index. With the field $\Psi$ and its antifield
$\Psi^*_i$, the antibracket between $A=\int[dZ]_ia_i$ and
$B=\int[dZ]_ib_i$ is
$$
(A,B)_{\hbox{\sixrm vector}}=\int \left(a_i{\larrowover\*\over\*\Psi}[dZ]_j
{\rarrowover\*\over\*\Psi^*_j}b_i
-a_i{\larrowover\*\over\*\Psi^*_j}[dZ]_j
{\rarrowover\*\over\*\Psi}b_i
\right)\punkt\eqn
$$
For the self-conjugate matter field $\Phi$,
$$
(A,B)_{\hbox{\sixrm matter}}=\int a_i{\larrowover\*\over\*\Phi}e_j[dZ]_j
{\rarrowover\*\over\*\Phi^\dagger}b_i\punkt\eqn
$$


\subsection\OperatorSS{Some useful operators}It was already observed
that, in order to write the equations of motion for the physical
fields (in the cohomology of $Q$), a triplet operator $\hat H_i$ with
dimension 2 
and ghost number $-1$ is needed. The
r\^ole of the operator is essentially to create a new (triplet) pure
spinor superfield which in the minimal picture would have the
auxiliary field $H_i$ as its $\lambda$- and $\theta$-independent
component.  In ref. [\CederwallKarlssonBI], similar operators were
formed (in the context of maximally supersymmetric SYM) corresponding
to various physical fields). 

The first observation is that there are other nilpotent operators than
$Q$. Also the expressions $q_i=(\lambda \sigma_iD)$ are nilpotent modulo
the pure spinor constraint. They can be extended to
$$
Q_i=(\lambda \sigma_iD)+(d\lb\sigma_i{\*\over\*\lb})\eqn
$$
in order to act non-trivially in the non-minimal sector.
Then, $\{Q,Q_i\}=0$, $\{Q_i,Q_j\}=0$.

A commonly used type of operator in pure spinor field (and string)
theory is the $b$-operator. It has the property
$$
\{Q,b\}=-\square\komma\eqn
$$
and clearly has ghost number $-1$ and dimension 2.
An explicit form of $b$ is
$$
\eqalign{
b&=\fr2(\lambda\bl)^{-1}(\bl\gamma^aD)\*_a\cr
&-\fr4(\lambda\bl)^{-2}(\bl\gamma^a\sigma^id\bl)
                        \left(N_i\*_a-\fr8(D\gamma_a\sigma_iD)\right)\cr
&\qquad  -\fr{16}(\lambda\bl)^{-2}(\bl\gamma^{abc}d\bl)
           \left(N_{ab}\*_c-\fr{24}(D\gamma_{abc}D)\right)\cr
&-\fr{32}(\lambda\bl)^{-3}((\bl d\bl^2)^a\gamma^bD)N_{ab}
       -\fr{16}(\lambda\bl)^{-3}((\bl d\bl^2)^iD)N_i\cr
&-\fr{64}(\lambda\bl)^{-4}(\bl d\bl^3)^{abi}N_{ab}N_i
    -\fr{64}(\lambda\bl)^{-4}(\bl d\bl^3)^{ij}N_iN_j\cr
}\Eqn\BOperator
$$
(see appendix \GammaMatrixAppendix\ for notation for antisymmetric
products of spinors).

The operators $Q_i$ and $b$ will not be used further in the present
paper, but will be of use when gauge fixing is considered. We turn to
the construction of the operator $\hat H_i$.
The precise criterion on $\hat H_i$ is that $\{Q,\hat H_i\}=0$ modulo the shift
transformations of eq. (\AntiFieldShift). This is satisfied by the
operators
$$
\eqalign{
\hat H_i&=(\lambda\bl)^{-2}(\bl\gamma^a\sigma_id\bl)\*_a
-\fr2(\lambda\bl)^{-3}(\bl d\bl^2)_i^\alpha D_\alpha\cr
&+(\lambda\bl)^{-4}\left[
\fr4(\bl d\bl^3)_{ij}N^j+\fr8(\bl d\bl^3)_{abi}N^{ab}
\right]\punkt\cr}
\Eqn\HOperator
$$
Note that the minimal representative for the auxiliary field
cohomology is at $\Psi\sim\lambda\theta^3$, a component yielding a
non-vanishing regularised integral $\int\Omega_i\wedge\Psi\sim H_i$. It
would seem that $\hat H_i$ should contain three spinorial
derivatives\foot{In ref. [\CederwallNilssonSix] such
an operator was constructed using minimal pure spinor variables. It
had the drawback of not being
well-defined outside cohomology.}.
Instead it contains  terms with $Dd\bl^2$ and $d\bl^3$,
which in the
integral with regularisation according to eq.
(\Regulator) can be converted into
fermionic derivatives. The expression (\HOperator), being linear in derivatives,
follows the pattern of
similar operators constructed in ref. [\CederwallKarlssonBI].

The linearised equations of motion for $\Psi$, already subject to $Q\Psi=0$, can
now be written as $\hat H_i\Psi=0$. 


\subsection\SYMActionSS{SYM action}We are now ready to write down the
BV action for the SYM multiplet in $\Psi$ and its antifield
$\Psi^*_i$.
The linearised action is
$$
S_{0,\hbox{\sixrm vector}}
=\int[dZ]_i\hbox{Tr}\left(\Psi^*_iQ\Psi+\fr2\Psi \hat H_i\Psi\right)
\punkt\eqn
$$

It is somewhat easier to check the master equation by repeated
variations on the field and antifield.
The equations of motion following from the action are
$$
\eqalign{
&Q\Psi=0\komma\cr
&Q\Psi^*_i+\hat H_i\Psi=0\punkt
}\eqn
$$
If the second equation is seen as a condition on $\Psi$ (effectively,
the vanishing of the auxiliary fields), the first term is trivial in
the cohomology.
The consistency amounts to the nilpotency of the operator
$$
{\cal Q}=\left(\matrix{Q&0\cr
      \hat H_i&Q}\right)\komma\eqn
$$
acting on the vector $(\Psi,\Psi^*_i)^t$, again modulo shift symmetry
in the second entry.

\subsection\MatterActionSS{Matter action}The matter field is
self-conjugate, $Q\Phi=0$ puts the component fields on shell,
and it is straightforward to write down an action. Suppressing indices
for the representation of $\Phi$ under the gauge group,
$$
S_{0,\hbox{\sixrm matter}}=\fr2\int[dZ]_i\Phi^\dagger e_i Q\Phi\punkt\eqn
$$
Note that the shift transformation $\delta_\rho\Phi=\lambda^\dagger\rho$
leads to a change in the action
$$
\delta_\rho S_{0,\hbox{\sixrm matter}}
=\fr2\int[dZ]_i(\Phi^\dagger e_i\lambda^\dagger
Q\rho+\rho^\dagger\lambda e_iQ\Phi)=0\komma\eqn
$$
where both terms vanish due to the property (\LambdaOmega) of the
integration measure.

\subsection\InteractionSS{Interactions}Interactions are introduced by
``covariantisation'' of the linearised action, so that the ``field
strength'' $Q\Psi$ is replaced by $Q\Psi+\Psi^2$. At the same time,
$Q\Phi\rightarrow(Q+\Psi\cdot)\Phi$ (the dot denoting action of the
gauge algebra in the representation of $\Phi$). This gives the complete action
for SYM coupled to matter:
$$
S=\int[dZ_i]\hbox{Tr}\left(\Psi^*_i(Q\Psi+\Psi^2)+\fr2\Psi
\hat H_i\Psi\right)
+\fr2\int[dZ]_i\Phi^\dagger e_i(Q+\Psi\cdot)\Phi\punkt\Eqn\FullAction
$$
Note that although the component interactions, both between gauge
fields and between scalars in the matter multiplets, include quartic
terms, the present formalism only gives 3-point couplings. The quartic
terms will arise when the superfield identities are solved, \ie, when
non-physical components are eliminated.
This is a typical feature of the pure spinor superfield 
formalism, and the present behaviour mirrors that of maximally
supersymmetric SYM. Even more drastic reduction of the order of the
interactions are seen in the actions for BLG and ABJM models
[\CederwallBLG,\CederwallABJM,\CederwallThreeConf],
in the Born--Infeld deformation of $D=10$ SYM
[\CederwallKarlssonBI,\ChangDeformationsI,\ChangDeformationsII], and 
in $D=11$ supergravity [\PureSGI,\PureSGII].

The equations of motion following from the action (\FullAction) are
$$
\eqalign{
(S,\Psi)&=Q\Psi+\Psi^2=0\komma\cr
(S,\Psi^*_i)&=Q\Psi^*_i+[\Psi,\Psi^*_i]+\hat H_i\Psi
        -\fr2\Phi^\dagger\circ e_i\Phi=0\komma\cr
(S,\Phi)&=(Q+\Psi\cdot)\Phi=0\komma\cr
}\Eqn\InteractionsEq
$$
where ``$\circ$'' is shorthand for formation of the adjoint of the
gauge algebra, and $[\cdot,\cdot]$ denotes adjoint action.
Note that gauge field interactions are introduced by deformation
(covariantisation) of the cohomology, while the matter current
back-reacts on the SYM fields through a deformation of the condition on
the auxiliary fields (the current multiplet).


When checking that the master equation $(S,S)=0$ is satisfied, one
finds that it relies on $\{Q,\hat H_i\}=0$, but also on the
distributivity of $\hat H_i$,
$\hat H_i(\Psi^2)=\hat H_i\Psi\Psi-\Psi\hat H_i\Psi$. This holds
thanks to the linearity of $\hat H_i$ in derivatives.


Concerning other modifications, it should be straightforward to apply
the method of ref. [\CederwallKarlssonBI] in order to write possible
higher-derivative interaction terms. Then there is no need to deform
the gauge transformations, which should mean that the first equation
in (\InteractionsEq) can be left unchanged, \ie, additional terms do
not contain the antifield. All new interaction then comes through
modification of the on-shell condition
$\hat H_i\Psi\sim\hbox{trivial}+J_i$.

%
%

\section\Conclusions{Conclusions}We have presented a classical
Batalin--Vilkovisky action for chiral $D=6$ SYM
theory. The gauge multiplet is not maximally supersymmetric, and
consequently its equations of motion are not implied by the cohomology
of the pure spinor superspace BRST operator. The hypermultiplet, on
the other hand, is maximally supersymmetric, and supersymmetric action
requires this kind of action.
The construction may stand model for superspace formulations of other
half-maximal models, like \eg\ $D=10$, $N=1$ supergravity.

The quantum theory has not been addressed.
It seems likely that models of the present type could serve as an
arena for the investigation of a complete and systematic gauge fixing
procedure for theories formulated on pure spinor superspace.
At the
present level of understanding, the constraint ``$b\Psi=0$''
reproduces Lorentz gauge and other appropriate conditions on
antifields, but how it is to be incorporated in a
systematic way in the BV formalism, using a gauge fixing fermion,
remains to be investigated. Simplifications may occur when fields and
antifields are separated.
This will be the subject of future work.

\appendix\GammaMatrixAppendix{Some spinor relations}When constructing
the operators of negative ghost number, 
completely antisymmetric products of spinors
are needed. All terms in $b$ and
$\hat H_i$ contain $\bl_{[\alpha_1}d\bl_{\alpha_2}\ldots
d\bl_{\alpha_p]}$.
The complete list of antisymmetrised spinors up to fourth order
is given in Figure
\AntiSymmSpinorFig.

\epsfxsize=5cm
\Figure\AntiSymmSpinorFig{\epsffile{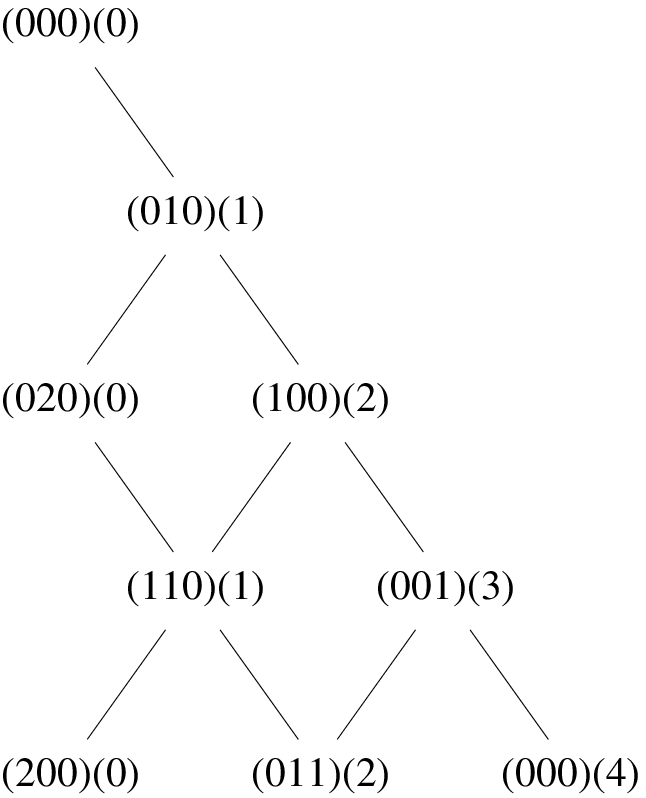}}{Irreducible
representations in antisymmetric products of spinors.}

The general antisymmetric bilinear Fierz identity, conveniently
expressed with the help of a fermionic spinor $s_\alpha$, is
$$
s_\alpha s_\beta=\fr8(\gamma_a\sigma_i)_{\alpha\beta}(s\gamma^a\sigma^i s)
+\fr{96}(\gamma_{abc})_{\alpha\beta}(s\gamma^{abc}s)\punkt\eqn
$$
expressing $\wedge^2(010)(1)=(100)(2)\oplus(020)(0)$.
At third order, $\wedge^3(010)(1)=(110)(1)\oplus(001)(3)$, represented
by
$$
\eqalign{
(s^3)^a_\alpha&=(\sigma_is)_\alpha(s\gamma^a\sigma^is)\komma\cr
(s^3)^{i\alpha}&=(\gamma_as)^\alpha(s\gamma^a\sigma^is)\punkt\cr
}\eqn
$$
One also has the identity
$$
(\gamma_{bc}s)_\alpha(s\gamma^{abc}s)=-4(\sigma_is)_\alpha(s\gamma^a\sigma^is)
\punkt\eqn
$$
At fourth order,
$\wedge^4(010)(1)=(200)(0)\oplus(011)(2)\oplus(000)(4)$. They can be
constructed from the cubic or quadratic expressions as
$$
\eqalign{
(s^4)^{ab}&=(s\gamma^a(s^3)^b)=(s\gamma^a\sigma_is)(s\gamma^b\sigma^is)\komma\cr
(s^4)^{abi}&=(s\gamma^a\sigma^i(s^3)^b)
        =-\epsilon^{ijk}(s\gamma^a\sigma_js)(s\gamma^b\sigma_ks)\komma\cr
(s^4)^{ij}&=(s\sigma^i(s^3)^j)=(s\gamma_a\sigma^is)(s\gamma^a\sigma^js)\punkt\cr
}\eqn
$$
A dependent expression for $(011)(2)$ is
$$
(s\gamma^{ab}(s^3)^i)=(s\gamma^{abc}s)(s\gamma_c\sigma^is)
=-2(s^4)^{abi}\punkt\eqn
$$
Since the dimension of the spinor module is 8, higher antisymmetric
products follow. The construction of the measure relies on
$$
\Omega_i=(\lambda\bl)^{-1}(\bl(d\lambda^5)_i)\eqn
$$
with $(d\lambda^5)^\alpha_i=(\sigma^jd\lambda)^\alpha
(d\lambda\gamma^a\sigma_jd\lambda)(d\lambda\gamma_a\sigma_id\lambda)$
in $(001)(3)$.

\acknowledgements{The author would like to thank Evgeny Ivanov for a
discussion inspiring me to continue on an old project, and Bengt
E.W. Nilsson for collaboration on that old project
[\CederwallNilssonSix]. I would also like to thank Evgeny Ivanov and
Ioseph Buchbinder for a discussion on 
gauge fixing.}

\refout

\end